\documentclass[aps,prc,preprint,showpacs,amsmath,amssymb,nofootinbib]{revtex4}
\usepackage {lscape}
\usepackage{graphicx}
\usepackage{amsmath}
\usepackage[export]{adjustbox}
\usepackage {color}
\usepackage{longtable}
\usepackage[bookmarks=true]{hyperref}
\usepackage{color}

  \hypersetup{
   colorlinks=true,
   pdfborder={0 0 0},
   citecolor=blue,
   linkcolor=blue,
   urlcolor=blue
 }

\begin{document}

%\title{{\Large Deformed nuclei with $ab-initio$ approaches and  $BE(2)$ values }}
\title{{\Large $Ab~initio$ description of collectivity for $sd$ shell nuclei}}
%\title{{\Large $BE(2)$ values with $ab~initio$ approaches }}% Force line breaks with \\
\author{\large A. Saxena$^1$}
%%\email{archanasaxena777@gmail.com}
\author{\large A. Kumar$^1$}
\author{\large V. Kumar$^2$}
\author{\large P. C. Srivastava$^1$}
\author{\large T. Suzuki$^{3,4}$}

\affiliation{$^1$Department of Physics, Indian Institute of Technology, Roorkee - 247667, INDIA}
\affiliation{$^2$Department of Physics, Central University of Kashmir, Ganderbal - 190004, INDIA}
\affiliation{$^3$Department of Physics, College of Humanities and Science, Nihon Univerity, Sakurajosui 3, Setagaya-ku, Tokyo 156-8550, JAPAN}
\affiliation{$^4$ National Astronomical Observatory of Japan, Osawa 2, Mitaka, Tokyo 181-8588, JAPAN  
 }

%%%%%%%%%%%%%%%%%%%%%%%%%%%%%%%%%%%%%%%%%

\begin{abstract}
In the present work, we have reported shell model results for open shell  nuclei Ne, Mg and Si isotopes with $10 \leq N \leq 20$
in  $sd$-shell model space.
We have performed calculations in $sd$ shell with two $ab~initio$ approaches:
in-medium similarity renormalization group (IM-SRG) and coupled-cluster  (CC) theory. We have also performed 
calculations with phenomenological USDB interaction and  chiral effective field theory based CEFT interaction.
The results for rotational spectra  and $B(E2;2_1^+\rightarrow 0_1^+)$ transitions are reported for even-mass isotopes. 
The  IM-SRG and CC results are in reasonable agreement with the experimental data  except at $N$ =20. This demonstrates a validity of 
$ab~initio$ description of deformation for doubly open-shell nuclei for $sd$ shell. To see the importance of $pf$ orbitals,
we have also compared our results with SDPF-MU interaction by taking account of $2p-2h$ and $4p-4h$  configurations in  $sd$-$pf$-shell model space.
\end{abstract}
\pacs{21.60.Cs -  shell model}

\maketitle
\section{Introduction}

The structure and properties of nuclei significantly  change once we move towards  drip-lines \cite{Tsunoda,32Mg}.
The ``island of inversion" of nuclei around A = 30  has been a subject of several experimental and theoretical studies.
There are different portals of island of inversion,  for example,
they may occurs at $N=8$, $N=20$, $N=28$, $N=40$ and $N=50$.
To decide structure of these nuclei, intruder orbitals  also become important. 
Recently there are several experimental  groups which involve to study structure of these nuclei, 
in particular the ``island of inversion" region of light nuclei with $N\sim20$.
Thus, from theoretical side apart from naive shell model it is important to study these nuclei
with $ab~initio$ approaches \cite{stroberg,jansen}.  As it is now possible to study lower $sd$ shell nuclei using $ab~initio$ approaches due to 
advancement in the computational facility, it is challenging to test predictive power of $ab~initio$ calculations for doubly-open shell nuclei
 for the description of deformation in the medium-mass region.

In the present work our motivation is to test  the $ab~initio$ Hamiltonians 
to calculate the spectra and  $B(E2)$ transitions for the doubly-open $sd$ shell nuclei.
Previously, we have reported electromagnetic properties, Gamow-Teller (GT) strengths and spectroscopic strengths of $sd$ shell nuclei in Refs. \cite{archana_prc, archana_gt, pcs_prc}.

We organize present work as follows. In section 2, we present details about Hamiltonians for $ab ~ initio$ calculations.
In section 3, we present theoretical results along with experimental data wherever it
is available. Finally summary and conclusions are drawn in section 4.

\section{Formalism}
In our studies of neutron-rich Ne, Mg and Si isotopes, we have performed calculations in $sd$ space.  As it is well known that 
excitation to $pf$-shell is very important for nuclei  which belong to ``island of inversion" \cite{mg_brown}, thus, we have also
performed calculations in $sd$-$pf$ shell.

For the $sd$ shell,  we use the $ab~initio$  Hamiltonian derived from two modern $ab ~ initio$ approaches: IM-SRG
\cite{stroberg} and CCEI \cite{jansen,jan1}. We have also compared results with a phenomenological USDB effective 
interaction \cite{usdb}. For $sd-pf$ shell we have performed calculations with SDPF-MU intercation \cite{utsuno12}.
For the diagonalization of matrices we have used shell-model code  KSHELL \cite{KShell}.

 Using IM-SRG approach \cite{bogner} based on chiral two- and three-nucleon interactions, 
 Stroberg $\it {et ~ al.,}$ derived mass-dependent Hamiltonians for $sd$ shell nuclei \cite{stroberg}.
 The strategy is that we select a Hamiltonian in certain basis  so that the
energy states which strongly  differ in energy range in off diagonal matrix elements should be eliminated. 
After applying unitary transformation we get a final
Hamiltonian $H(s)$ from a initial Hamiltonian $H(0)$.

\begin{equation}
  H(s)=U^{\dag}(s)H(0)U(s)\,=H^{d}(s)+H^{od}(s).
\end{equation}

Here, s is the flow parameter. $H^{d}(s)$ is the diagonal part of the Hamiltonian and $H^{od}(s)$ is
the off diagonal part of the Hamiltonian.
 The flow of the Hamiltonian is obtained by differentiating Eq. (1) w.r.to `s'-

\begin{equation}
\frac{dH(s)}{ds}=[\eta(s),H(s)],
\end{equation}

 where $\eta(s)$ is the anti-Hermitian generator of unitary transformation, 
\begin{equation}
\eta(s)=\frac{dU(s)}{ds}U^{\dag}(s)=-\eta^{\dag}(s).
\end{equation}

Eq. (2) is  known as the flow equation for the Hamiltonian.  
The off-diagonal matrix 
 elements become zero as $s\rightarrow\infty$ for appropriate value of $\eta(s)$.
Here $sd$ valence space decouple from the core and higher shells as $s\rightarrow\infty$. Now we use  the resulting Hamiltonian
in the shell model calculations with $\hbar\omega$=24 MeV.  Further details about parameters are given in ref. \cite{stroberg}.

The Hamiltonian developed from the Coupled Cluster Effective Interaction approach is $A$- dependent and can be extended as-
\begin{equation}
{H_{CCEI}}=H_0^{A_c}+H_1^{A_c+1}+H_2^{A_c+2}+...  
\end{equation}
Here, $H_0^{A_c}$, $H_1^{A_c+1}$, and $H_2^{A_c+2}$ are called core,  one-body, and two-body cluster Hamiltonians, respectively.
This expansion is known as valance cluster expansion. Any operator can be expanded in the valence space 
 in the same way as the Hamiltonian for the shell model calculations. The Okubo-Lee-Suzuki (OLS) similarity transformation is used to calculate the two-body term. 
In this approach Hartree-Fock ground state in thirteen oscillator major shells with $\hbar\omega$=20 MeV is used.
 Recently, L. Huth et al. \cite{Huth} derived  a shell-model  interaction from effective field theory, which will be referred as CEFT. 

The $B(E2)$ values are  calculated with the formula:
\begin{equation}
       B(E2)= \frac{1}{2J_{i}+1}  
        \mid(J_f\mid\mid \sum_{i} e_{i}r_{i}^2 { Y_{2}}(\theta_{i},\phi_{i})\mid\mid\ J_i)\mid^2. 
\end{equation} 
Where, $J_i$ and $J_f$ are the initial and final state spins, respectively. 
The  $B(E2)$ values are calculated with the effective charges $e_{p}$=1.35e and $e_{n}$=0.35e.
%Previously shell model results of quadrupole moments for $sd$ shell nuclei reported in ref. \cite{brown_qm}.

\begin{figure*}
\begin{center}
\includegraphics[width=7.7cm,height=7cm,clip]{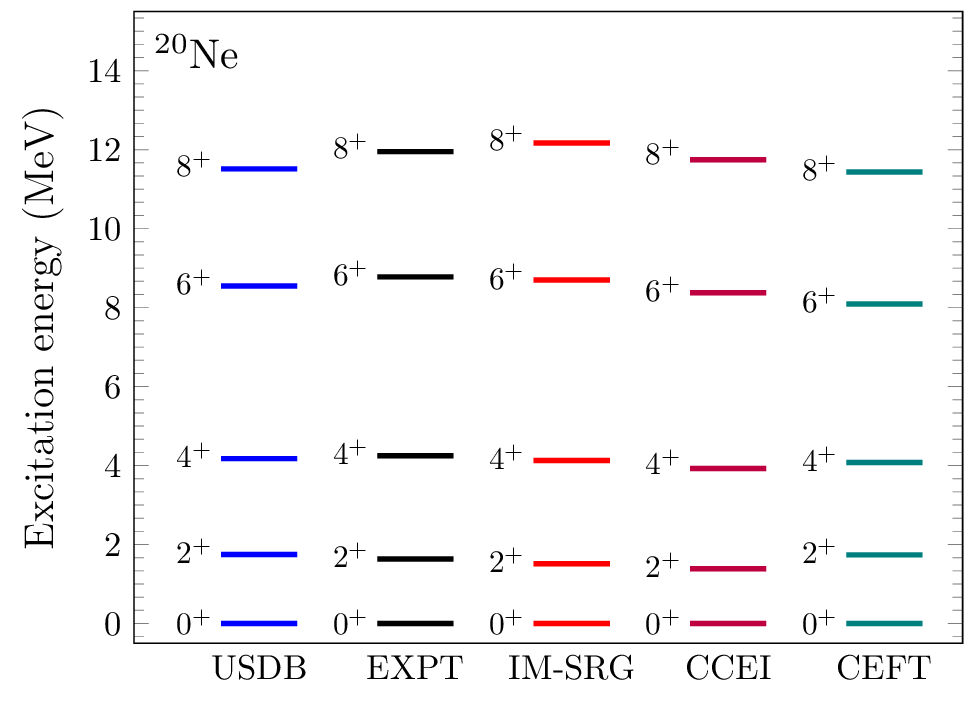}
\includegraphics[width=7.7cm,height=7cm,clip]{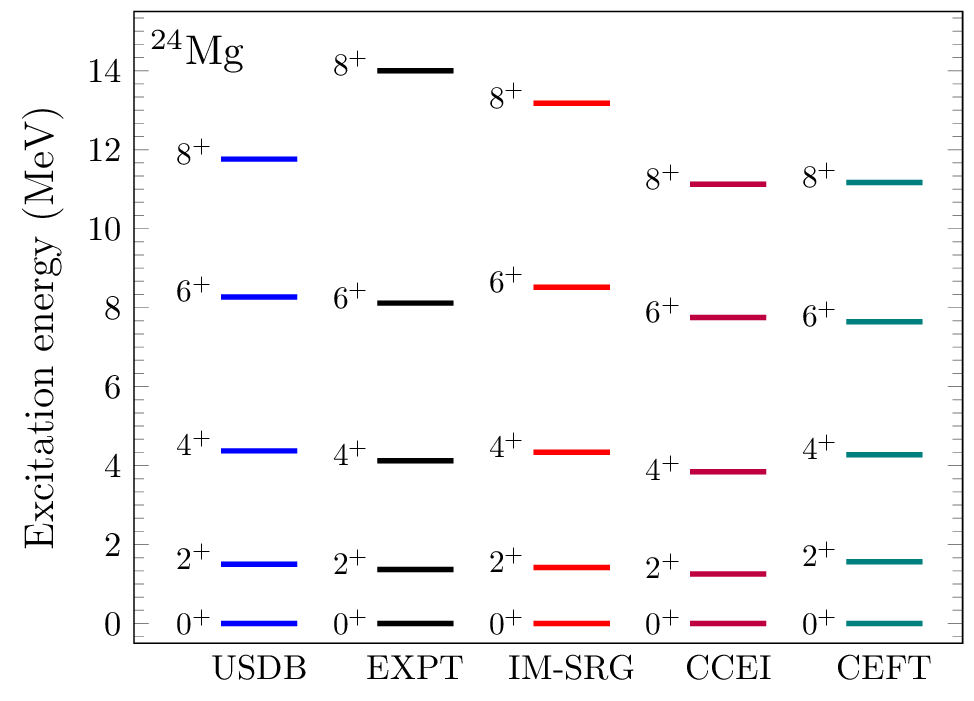}
\includegraphics[width=7.7cm,height=7cm,clip]{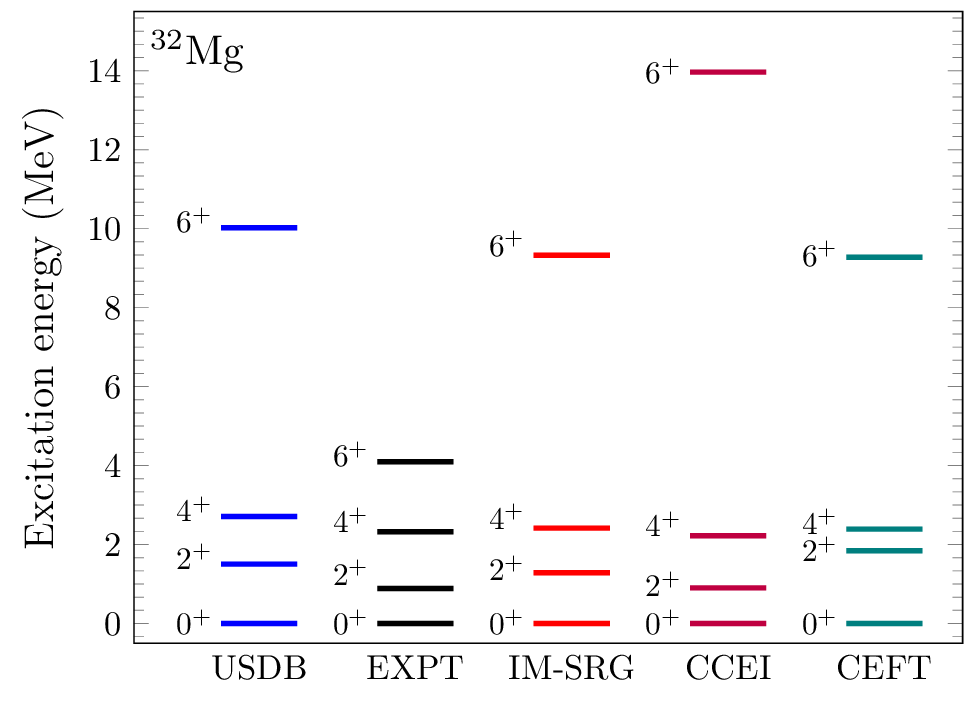}
\includegraphics[width=7.7cm,height=7cm,clip]{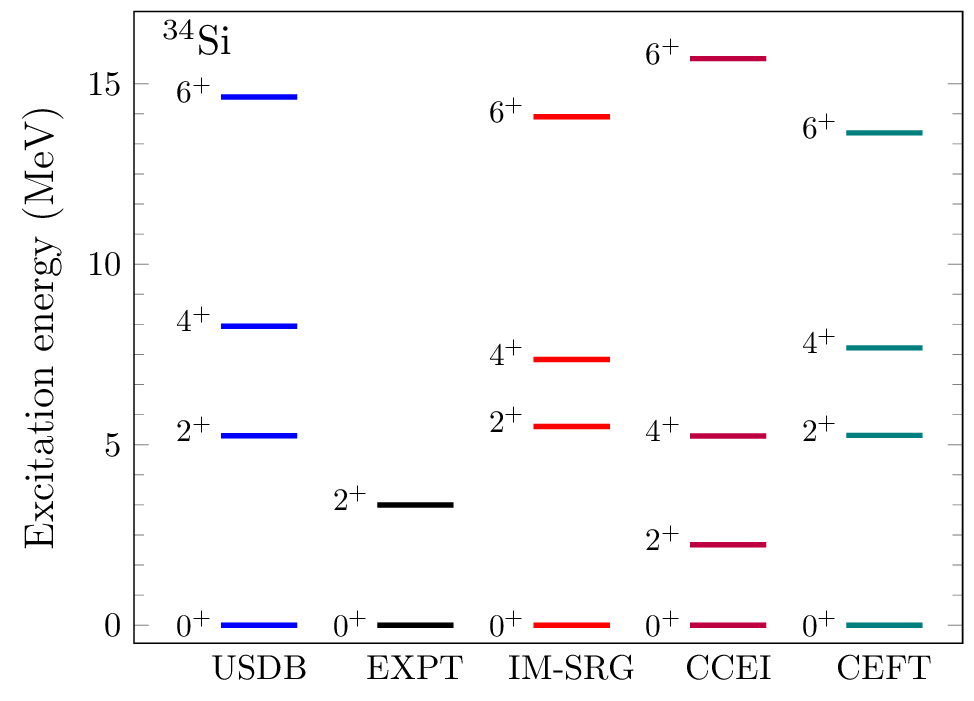}
\caption{\label{spectra}The energy spectra of $^{20}$Ne,$^{24}$Mg, $^{32}$Mg and $^{34}$Si using USDB, IM-SRG, CCEI and CEFT interactions.}
\end{center}
\end{figure*}

\section{Results and Discussions}

The experimental measurement of  $B(E2)$ values and nuclear moments are important to predict
existence of the ``island of inversion". 
The nuclei lying in the ``island of inversion" region show a drastic enhancement of quadrupole collectivity compared 
to neighboring nuclei from these experimental results.

\begin{figure*}
%\begin{center}
\includegraphics[width=7.7cm,height=7cm,clip]{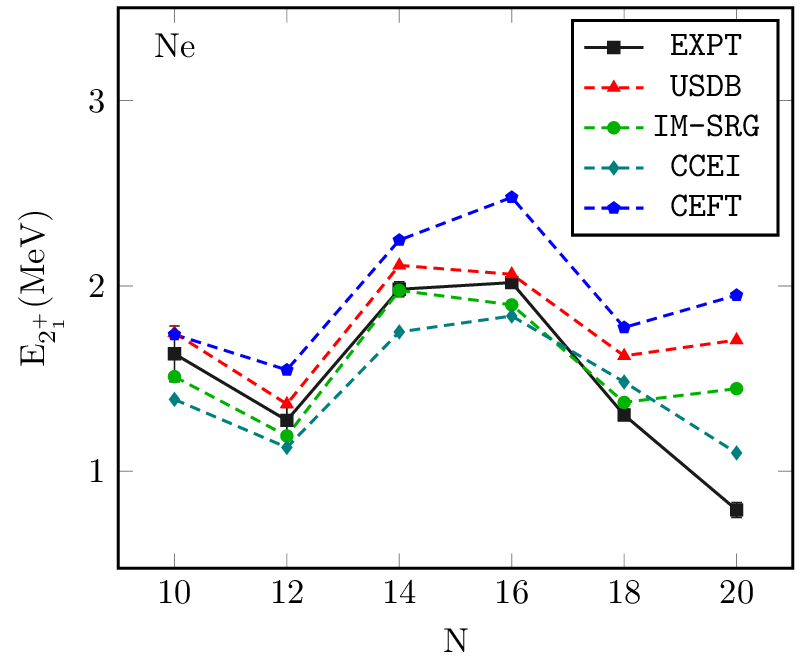}
\includegraphics[width=7.7cm,height=7cm,clip]{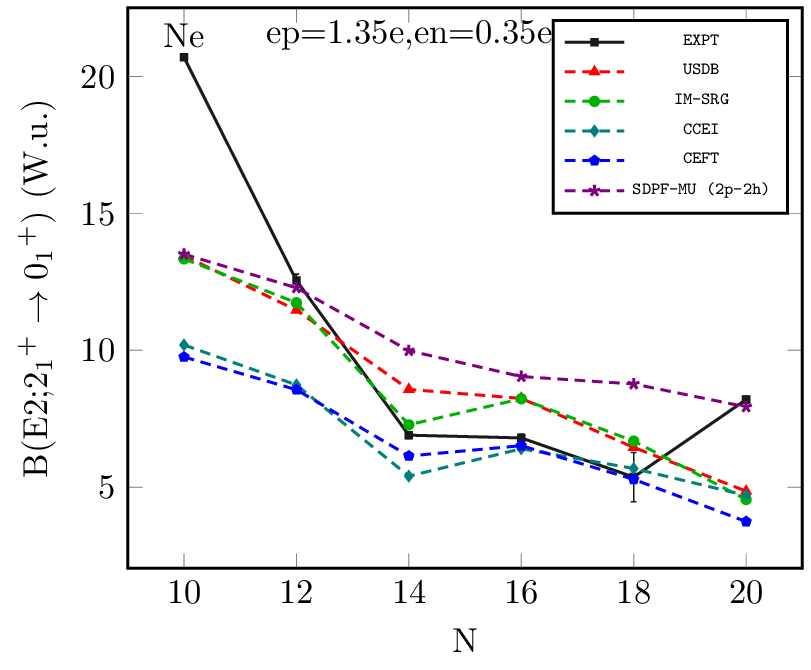}
\includegraphics[width=7.7cm,height=7cm,clip]{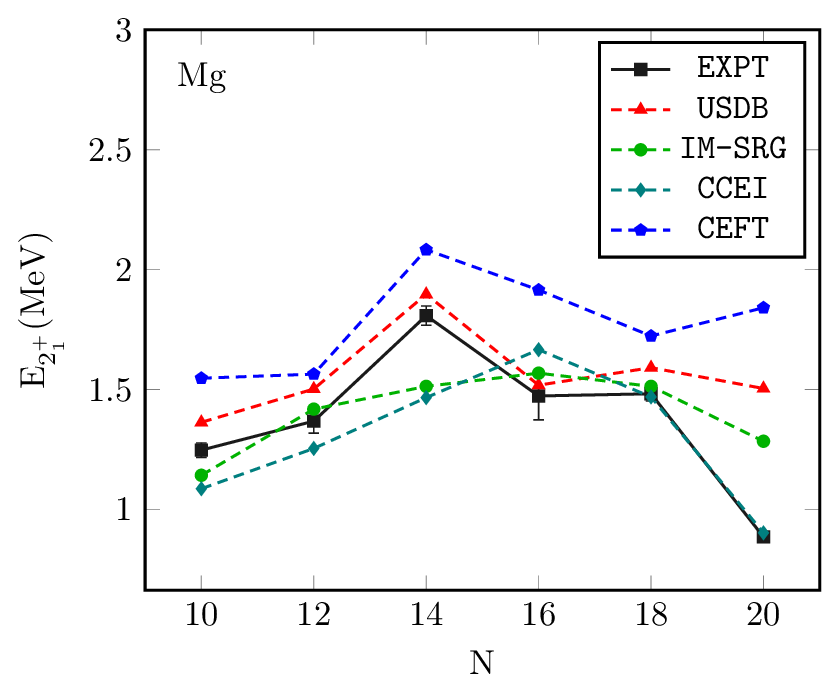}
\includegraphics[width=7.7cm,height=7cm,clip]{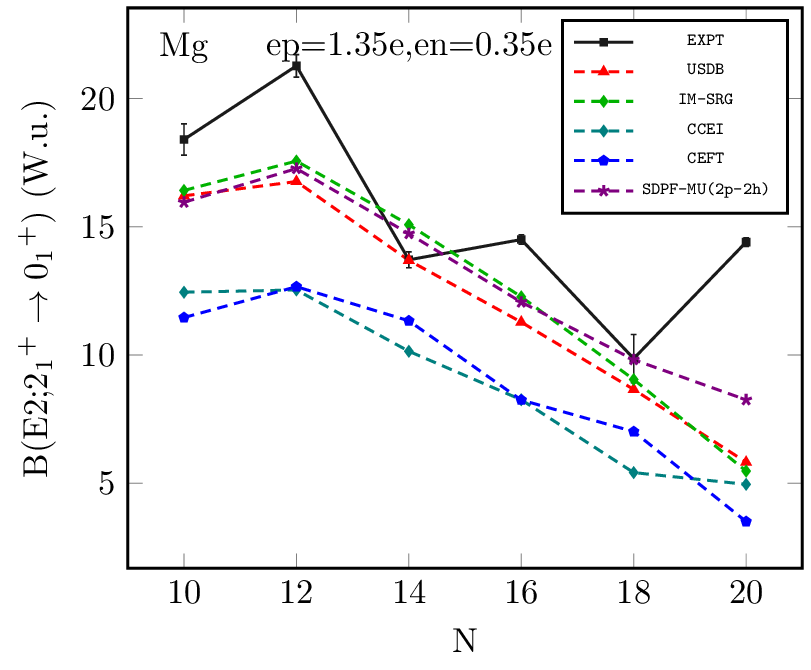}
\includegraphics[width=7.7cm,height=7cm,clip]{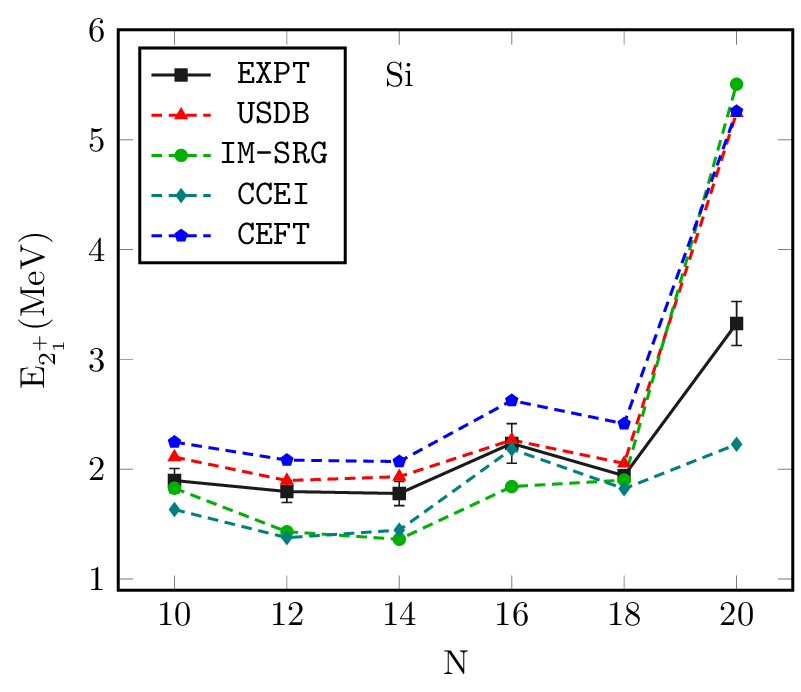}
\includegraphics[width=7.7cm,height=7cm,clip]{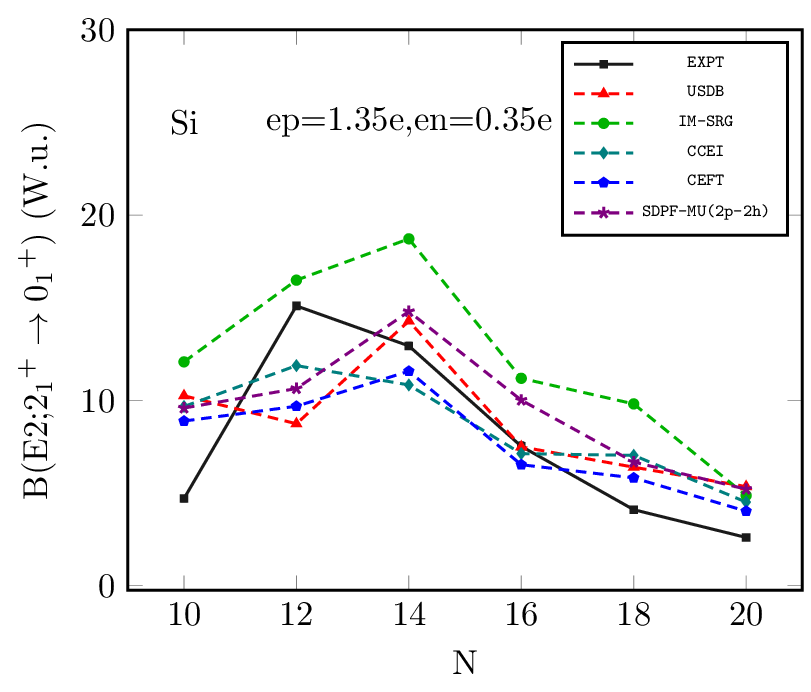}
\caption{\label{2+_spectra}The energy of $2_1^+$ and $B(E2;2_1^+\rightarrow 0_1^+)$  values of Ne, Mg and Si isotopes.}
%\end{center}
\end{figure*}

$^{20}$Ne  provides a good example of rotational spectra \cite{doubly_open} in the lower $sd$ shell. The comparison of 
rotational energy levels for $^{20}$Ne  obtained by USDB, IM-SRG, CCEI, and CEFT interactions  is shown in the Fig. \ref{spectra}.
 The rotational spectra $^{20}$Ne and $^{24}$Mg are reproduced in our $ab~initio$ calculations. The  $E_{2_1^+}$ and
$B(E2; 2_1^+\rightarrow 0_1^+)$ transitions using $ab~initio$ interactions and USDB interactions 
along with experimental data \cite{nndc, be2} for even Ne
isotopes with $N=10-20$ are shown in Fig. \ref{2+_spectra}. 
The USDB results for $E({2_1^+})$ are close to experimental data up to $N=16$ but above this results are deviating.
The IM-SRG results are  the best and close to the experimental data from $N=10$ to $N=18$.  At $N=20$,  only CCEI shows the same pattern as the experimental data, while results of all the other interactions 
go upward deviating from the experiment.  Experimentally, $N=18$ shows less collectivity in comparison to $N=16$ and an enhancement in collectivity 
 is seen at $N=20$.
 The $B(E2)$ values obtained are not satisfactory especially at $N=20$ for Ne and Mg isotopes.
   For all the interactions the collectivity is decreasing from $N=18$ to $N=20$  in contrast to the experiment.
From the literature the $N=20$ lies on the boundary of ``island of inversion" \cite{28Ne} and $0\hbar\omega$ shell model 
calculations are not able to reproduce the enhancement in collectivity at $N=20$. 

\begin{figure*}
\includegraphics[width=7.7cm,height=7cm,clip]{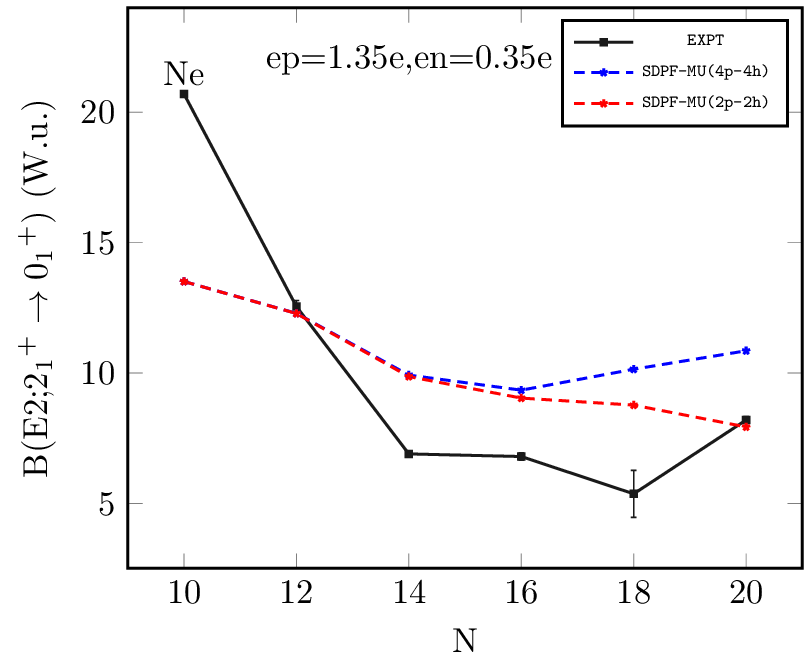}
\includegraphics[width=7.7cm,height=7cm,clip]{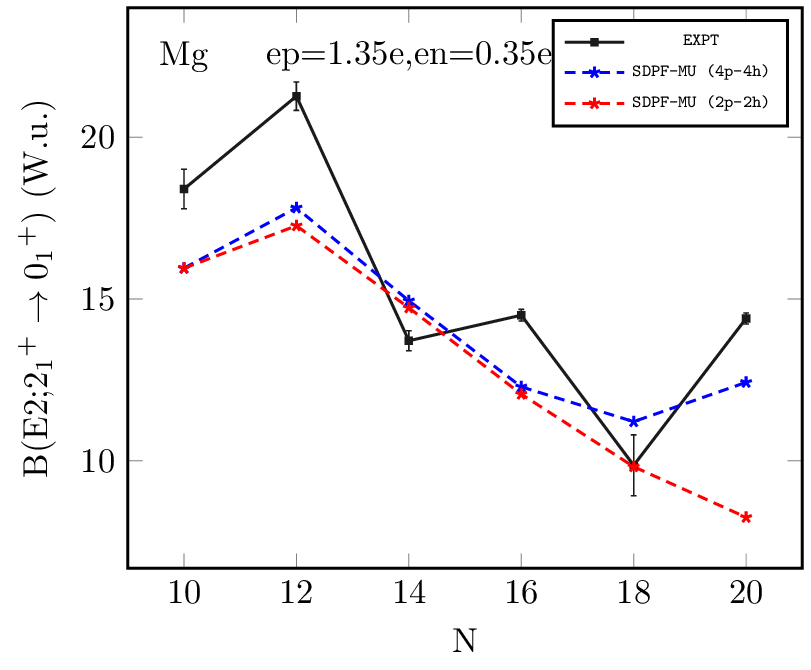}
\begin{center}
\includegraphics[width=7.7cm,height=7cm,clip]{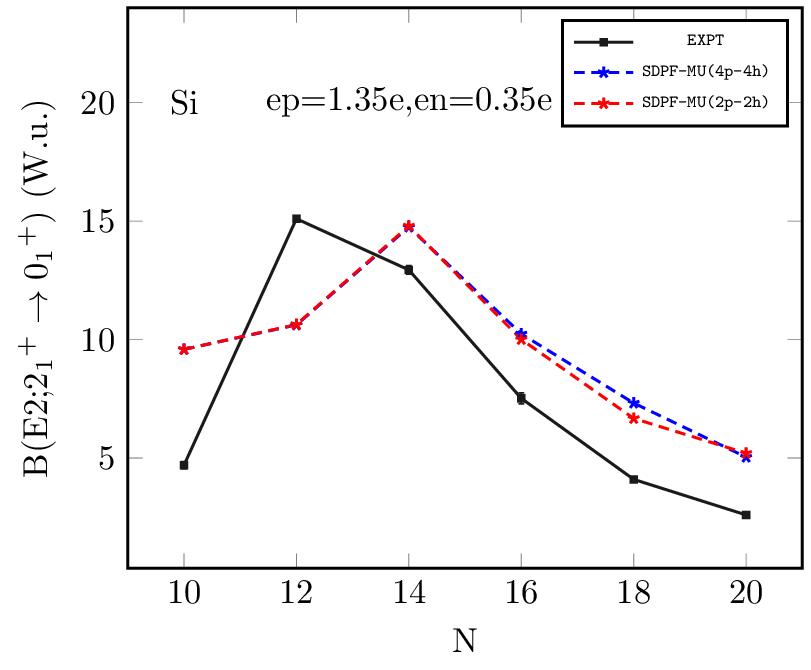}
\end{center}
\caption{\label{be2_sdpf}Comparison between calculated and experimental $B(E2;2_1^+\rightarrow 0_1^+)$  values of Ne, Mg and Si isotopes
for $sd-pf$ shell with $2p-2h$ and $4p-4h$ excitations.}
\end{figure*}

\begin{figure*}
%\begin{center}
\includegraphics[width=7.7cm,height=7cm,clip]{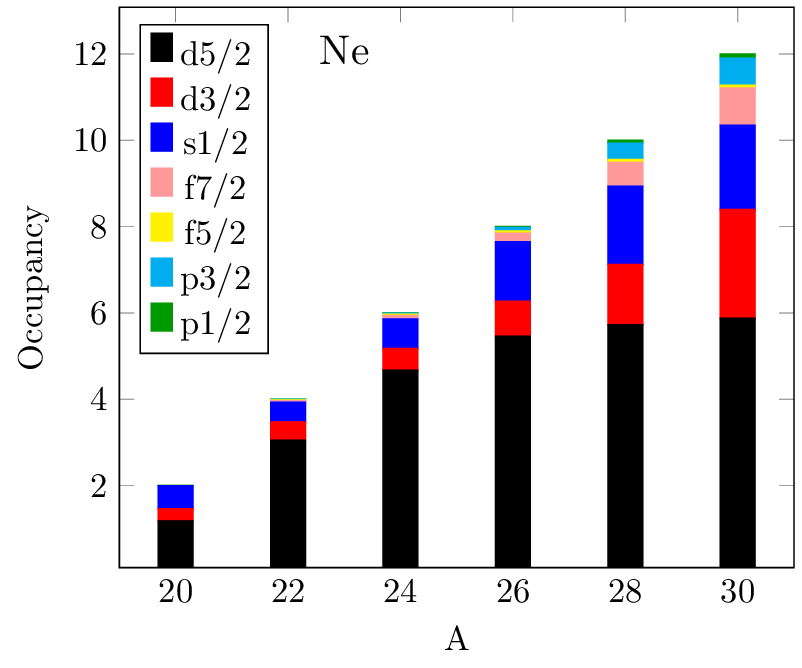}
\includegraphics[width=7.7cm,height=7cm,clip]{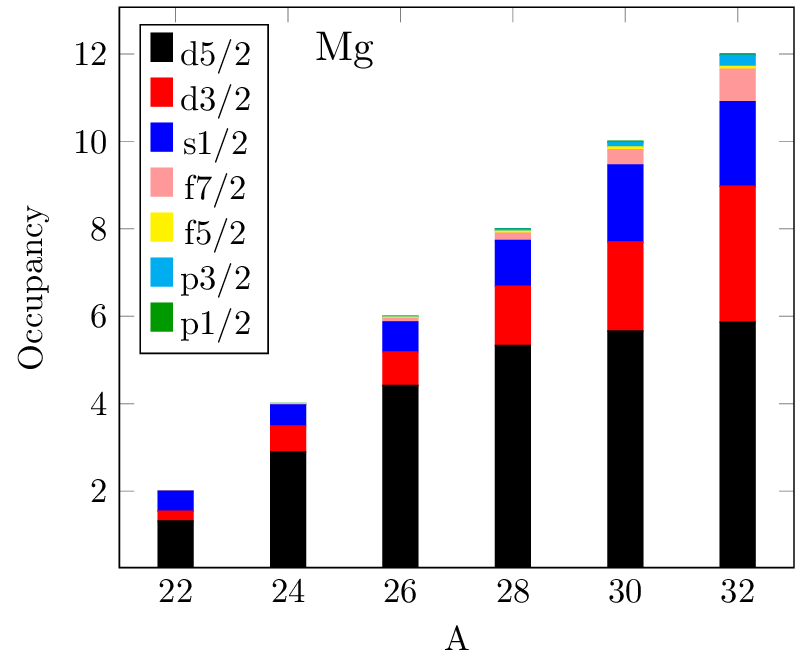}
\begin{center}
\includegraphics[width=7.7cm,height=7cm,clip]{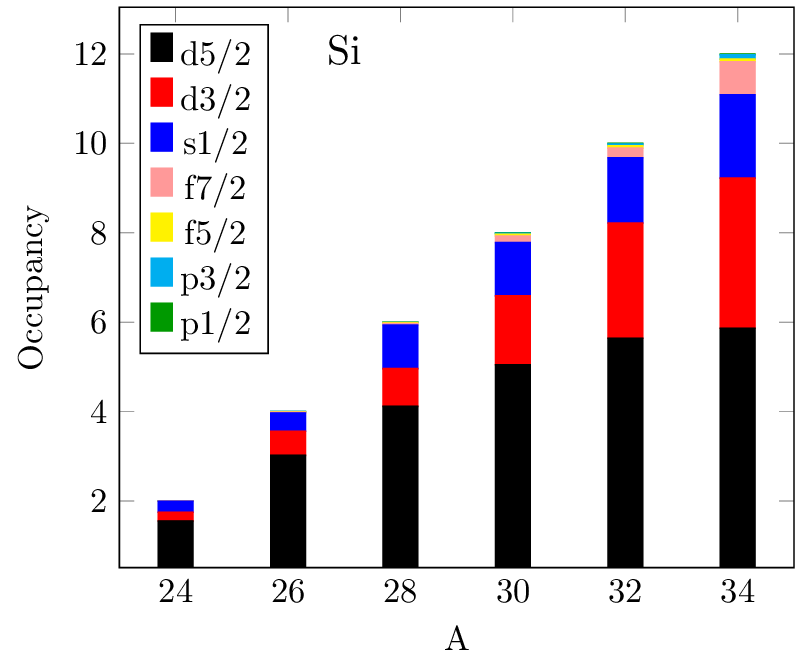}
\end{center}
\caption{\label{ocuupancy} The neutron occupancy of $2_1^+$ state  for Ne, Mg and Si isotopes with SDPF-MU interaction  with $2p-2h$ excitations .}
%\end{center}
\end{figure*}

To see the importance of neutron excitations from $sd$ to $pf$ shell, we have  shown results  with $2p-2h$ and $4p-4h$ excitations
in Fig. \ref{be2_sdpf}.  The results with $4p-4h$ excitations show the same trends as in the experiment : there is an increase of
$B(E2;2_1^+\rightarrow 0_1^+)$ from $N=18$ to $N=20$  though not enough to reproduce the experimental data. The increase in the occupancy of $pf$ orbitals are also
 confirmed from Fig. \ref{ocuupancy}.

%%%%%%%%%%%%%%
For the Mg isotopes, the case of $^{32}$Mg is very interesting. Its g.s. is deformed and it shows rotational spectrum. Up to $4^+$, the CCEI
results are very close to the experimental data but $6^+$ state is very high in energy.
Also, for $^{24}$Mg, the rotational spectra is well reproduced by $ab~initio$ interactions up to $6^+$.
In Fig. \ref{2+_spectra}, the energy of $2_1^+$ state is near to  the experimental data with IM-SRG effective interaction except at $N=14$ and $N=20$. 
The CCEI results are also good except at $N=14$. At $N=20$ the energy of $2_1^+$ state is correctly given by  CCEI but $B(E2)$ value is far 
from the experimental data. This shows there is a problem with the wavefunction, that is,
there is a large configuration mixing in this wavefunction.
Here, with $ab~initio$ interactions it is not possible to show collectivity at $N=20$.
The shell model results  with $2p-2h$ and $4p-4h$ exciations for SDPF-MU interaction are shown  
in Fig. \ref{be2_sdpf}. They show smooth decrease
in $B(E2)$ values from $N=12-16$ for both $2p-2h$ and $4p-4h$ excitations, however, 
the results of $4p-4h$  show the same trends as in the experiment at $N=20$ : there is an increase of
$B(E2;2_1^+\rightarrow 0_1^+)$ from $N=18$ to $N=20$.

%%%%%%%%%%%%%Si 
In the case of $^{34}$Si isotope none of the interaction explain properly the  spectra.
For Si isotopes the trend of energy of $2_1^+$ state for $N=10$ to $N=18$ isotopes are well predicted by CEFT 
and USDB interactions. The $B(E2;2_1^+\rightarrow 0_1^+)$ trend from $N=12$ to $N=18$ are showing 
reasonable agreement with  the experiment for all the interactions.
The $B(E2;2_1^+\rightarrow 0_1^+)$ values for  the case with $2p-2h$ and $4p-4h$ excitations  show similar results.

In the present calculations the gap between $d_{3/2}$ orbital and $fp$ shell is large for the SDPF-MU interaction compared with the interaction in Ref. \cite{Tsunoda}, where the neutron ESPE's of $pf$-shell are very close to those $sd$-shell; the difference between $f_{7/2}$ and $d_{3/2}$ is as small as about 2 MeV for Z=12 (N=20).
Therefore, even if we allow $4p-4h$ excitation from $sd$ to $pf$ shell, the occupancy of $fp$-shell are  around 2.23 in $^{32}$Mg. 
%Thus our results might be improved if we reduce gap between $d_{3/2}$ and $fp$ shells for SDPF-MU interaction. 
As we can see in Ref. \cite{Tsunoda}, where energies and $B(E2)$ are well reproduced up to $N=20$, the occupancy of $fp$-shell becomes as large as 3.5 for $^{32}$Mg.

%The neutron ESPE's of $pf$-shell are very close to those $sd$-shell; the difference between $f_{7/2}$ and $d_{3/2}$ is as small as about 2 MeV for Z=12 (N=20).
Thus our results might be improved if we reduce the gap between $sd$ and $fp$ shells for the SPDF-MU interaction.

\section{Summary and conclusions}

In the present work, we have performed shell model calculations  for open shell nuclei with  $10 \leq N \leq 20$ for Ne, Mg and Si
isotopes in the $sd$ and $sd-pf$ spaces. For $sd$ shell we have taken two $ab~initio$ interactions, IM-SRG and CCEI.
We have also performed calculations with phenomenological USDB and interaction based on chiral effective field theory. 
The degree of freedom of $sd$ and $pf$ shells are essential for the nuclei close to ``island of inversion"
so, we have also reported results of $sd-pf$ shell with SDPF-MU interaction. 
The results of $ab~initio$ interactions  show reasonable agreement with the experimental data  except at $N$ =20. 
 For nuclei in the island of inversion such as $^{30}$Ne and $^{32}$Mg, the admixture of $pf$ shells is important to
 explain the lowering of the energies of 2$_1^+$ states and the enhancement of the $B(E2)$ values.  
present study will add more information to earlier theoretical $B(E2)$ values of  Ne, Mg and Si isotopes.

\section*{Acknowledgments}

AS acknowledges financial support from MHRD (Govt. of India) for her Ph.D. thesis work.
Vikas Kumar acknowledges partial support from CUK.

%\newpage

\end{document}